# The electron-phonon coupling constant and the Debye temperature in polyhydrides of thorium, hexadeuteride of yttrium, and metallic hydrogen phase III

Evgueni F. Talantsev[1,2,*,**]

[1]M.N. Mikheev Institute of Metal Physics, Ural Branch, Russian Academy of Sciences, 18, S. Kovalevskoy St., Ekaterinburg, 620108, Russia

[2]NANOTECH Centre, Ural Federal University, 19 Mira St., Ekaterinburg, 620002, Russia

*corresponding author

**corresponding author's E-mail: evgney.talantsev@imp.uran.ru

**Abstract**

Milestone experimental discovery of superconductivity above 200 K in highly-compressed sulphur hydride by Drozdov *et al* (*Nature* **525**, 73 (2015)) sparked experimental and theoretical investigations of metallic hydrides. Since then, a dozen of superconducting binary and ternary polyhydrides have been discovered. For instance, there are three superconducting polyhydrides of thorium: $Th_4H_{15}$, $ThH_9$, and $ThH_{10}$ and four polyhydrides of yttrium: $YH_4$, $YH_6$, $YH_7$, $YH_9$. In addition to binary and ternary hydrogen-based metallic compounds, recently Eremets *et al* (*arXiv*:2109.11104) reported on the metallization of hydrogen, which exhibits a phase transition into metallic hydrogen phase III at $P \geq 330$ GPa and $T \sim 200$ K. Here we analysed temperature-dependent resistance, $R(T)$, in polyhydrides of thorium, hexadeuteride of yttrium and in hydrogen phase III and deduced the Debye temperature, $T_\theta$, and the electron-phonon coupling constant, $\lambda_{e-ph}$, for these conductors. We found that *I-43d*-$Th_4H_{15}$ exhibits $\lambda_{e-ph} = 0.82$-$0.99$, which is in a very good agreement with experimental value of $\lambda_{e-ph} = 0.84$ deduced from heat capacity measurements (Miller *et al*, *Phys. Rev. B* **14**, 2795 (1976)). For $P6_3/mmc$-$ThH_9$ ($P = 170$ GPa) we deduced $\lambda_{e-ph}(170\text{ GPa}) = 1.46 \pm 0.01$, which is in a reasonable agreement with $\lambda_{e-ph}$ computed by first-principles calculations ( Semenok *et al. Materials Today* **33**, 36 (2020)). Deduced $\lambda_{e-ph}(170\text{ GPa}) = 1.70 \pm 0.04$ for $Fm$-$3m$-$ThH_{10}$ is in remarkable agreement with first-principles calculated $\lambda_{e\text{-ph}}(174 \text{ GPa}) = 1.75$ (Semenok *et al*, *Materials Today* **33**, 36 (2020)). Deduced $\lambda_{e\text{-ph}}(172 \text{ GPa}) = 1.90 \pm 0.02$ for *Im-3m*-YD$_6$ is also in excellent agreement with first-principles calculated $\lambda_{e\text{-ph}}(165 \text{ GPa}) = 1.80$ (Troyan *et al*, *Advanced Materials* **33**, 2006832 (2021)). And finally, we deduced $T_\theta(402 \text{ GPa}) = 727 \pm 6$ K for hydrogen phase III which implies that $\lambda_{e\text{-ph}}(402 \text{ GPa}) \leq 1.7$ in this metal.

**The electron-phonon coupling constant and the Debye temperature in polyhydrides of thorium, hexadeuteride of yttrium, and metallic hydrogen phase III**

## I. Introduction

In 1935 Wigner and Huntington proposed [1] that molecular hydrogen can undergo a first-order phase transition with a dissociation into metallic phase at high pressure. Their calculations showed that the transition should occur at pressure of $P = 25$ GPa. Since then, as the experiment as the first-principles calculations dramatically increase the lower pressure boundary for the transition up to $P \gtrsim 450 \; GPa$ [2-5]. Surprisingly enough, hydrogen, the simplest element of the Mendeleev Table and the most abundant chemical element in the universe, exhibits perhaps the most complicated and rich phase diagram, which is not well explored to the date [2-8]. Detailed description of the status in this research field can be found elsewhere [6-8].

One of the most interesting hypotheses in regard of metallic hydrogen is its potential room-temperature superconductivity, which was first expressed by Ashcroft [9] and Ginzburg [10], while more practical idea of high-temperature superconductivity in hydrogen-rich compounds was expressed by Satterthwaite and Toepke [11]: "…There has been theoretical speculation [9] that metallic hydrogen might be a high-temperature superconductor, in part because of the very high Debye frequency of the proton lattice. With high concentrations of hydrogen in the metal hydrides one would expect lattice modes of high frequency and if there exists an attractive pairing interaction one might expect to find high-temperature superconductivity in these systems also." Based on this general idea, Satterthwaite and Toepke [11] synthesised the first superconducting superhydride $Th_4H_{15}$. This discovery was a part of much wider search for the superconductivity in polyhydrides [11,12]. For instance, Satterthwaite and Peterson [12] reported on the absence of the superconductivity above $T=$ 1.2 K in $ThH_2$, $(Th_{1/3}Zr_{2/3})H_{3.5}$, $VH_2$, $NbH_2$, $TaH$.

To date, there are three discovered superconducting polyhydrides of thorium, $I\bar{4}3d$-$Th_4H_{15}$ ($T_c \cong 8\ K$) [11], $P6_3/mmc$-$ThH_9$ ($T_c \cong 145\ K$) [13], and $Fm\bar{3}m$-$ThH_{10}$ ($T_c \cong 160\ K$) [13]. It should be noted that elemental thorium exhibits the superconducting transition temperature $T_c = 1.37\ K$ [14].

In 2015, more than four decades after the report by Satterthwaite and Toepke [11], Drozdov *et al* [15] reported on the discovery of the near-room temperature superconductivity in highly-compressed polyhydride of sulphur, $H_3S$, which is widely acknowledged [8,16,17] to be a confirmation for the electron-phonon pairing in hydrogen-rich materials. Further experimental and first-principles calculations studies leaded to the discovery of about a dozen of superconducting superhydrides with the superconducting transition temperature above 100 K [15,18-25] and several polyhydrides with $T_c$ below 100 K [26-33]. There is also a recent conjecture that the S orbitals in $H_3S$ play an active electronic role, in analogy to the O orbitals in high-$T_c$ cuprates [34]. Recent review of the first-principles calculations in the field of highly-compressed hydrides was given by Chen *et al* [35].

At the same time, attempts to discover room-temperature superconductivity in elemental metallic hydrogen have been never stopped and, recently, Eremets *et al* [36] reported on the metallization of hydrogen phase III at pressure of $P \geq 330$ GPa and temperature of $T \sim 200$ K.

Here, following general consensus [8,16,17] that the electron-phonon pairing is primary mechanism governs the superconductivity in highly-compressed polyhydrides, we deduced the Debye temperature, $T_θ$ (which is characteristic value of the parabolic approximation of full phonon spectrum) and the electron-phonon coupling constant, $λ_{e-ph}$, for:

1. all discovered to date polyhydrides of thorium, i.e. $I\bar{4}3d$-Th$_4$H$_{15}$, $P6_3/mmc$-ThH$_9$ and $Fm\bar{3}m$-ThH$_{10}$ phases (Sections 3.3-3.6);

2. *Im-3m*-YD$_6$ phase (Section 3.7);

3. hydrogen phase III compressed at $P$ = 402 GPa (Section 3.8).

Experimental data for $Fm\bar{3}m$-ThH$_{10}$ phase was kindly provided by Dr. D. V. Semenok and Prof. A. R. Oganov (Skolkovo Institute of Science and Technology), and data for *Im-3m*-YD$_6$ phase was kindly provided by Dr. A. G. Kvashnin, Dr. I. A. Troyan and Prof. A. R. Oganov (Skolkovo Institute of Science and Technology).

## 2. Utilized model

Debye temperature, $T_θ$, of the metallic conductor can be deduced as a free parameter from the fit of normal part of temperature dependent resistance, $R(T)$ (or resistivity, $ρ(T)$), to the Bloch-Grüneisen (BG) equation [37,38]:

$$R(T) = R_0 + A \times \left(\frac{T}{T_θ}\right)^5 \times \int_0^{\frac{T_θ}{T}} \frac{x^5}{(e^x-1)(1-e^{-x})} dx \quad (1)$$

where $R_0$, A and $T_θ$ are free-fitting parameters, and the former term is residual resistance appeared due to the conduction electrons scattering on the static defects of the lattice. There are obvious restrictions for free-fitting parameters in Eq. 1, that $R_0 \geq 0$ and $T_θ > 0$. In considered case of superconductors, Eq. 1 is only valid to fit the normal part of $R(T)$ and recently [39] we proposed to split full $R(T)$ curve in two part: the normal part (which is fitted by Eq. 1, i.e. for $T_c^{onset} < T$, where $T_c^{onset}$ is the onset of the superconducting transition), and the transition and zero resistance part is approximated by simplified equation proposed by Tihkham [40]:

$$R(T) = \frac{R(T_c^{onset})}{\left(I_0\left(F\times\left(1-\frac{T}{T_c^{onset}}\right)^{3/2}\right)\right)^2} \qquad (2)$$

where $F$ is free-fitting parameter and $I_0(x)$ is the zero-order modified Bessel function of the first kind. These two parts (Eqs. 1,2) are stitched at $T = T_c^{onset}$ using the Heaviside function, $\theta(x)$, and full fitting equation is [39]:

$$R(T) = R_0 + \theta(T_c^{onset} - T) \times \left(\frac{R(T_c^{onset})}{\left(I_0\left(F\times\left(1-\frac{T}{T_c^{onset}}\right)^{3/2}\right)\right)^2}\right) + \theta(T - T_c^{onset}) \times \left(R(T_c^{onset}) + A \times \left(\left(\frac{T}{T_\theta}\right)^5 \times \int_0^{\frac{T_\theta}{T}} \frac{x^5}{(e^x-1)(1-e^{-x})} dx - \left(\frac{T_c^{onset}}{T_\theta}\right)^5 \times \int_0^{\frac{T_\theta}{T_c^{onset}}} \frac{x^5}{(e^x-1)(1-e^{-x})} dx\right)\right) \qquad (3)$$

The Debye temperature, $T_\theta$, and the transition temperature, $T_c$, are linked through the McMillan equation [41], which was recently [42] simplified by excluding one parameter which cannot be deduced from real-world high-pressure experiments (see details in Ref. 42):

$$T_c = \left(\frac{1}{1.45}\right) \times T_\theta \times e^{-\left(\frac{1.04(1+\lambda_{e-ph})}{\lambda_{e-ph}-\mu^*(1+0.62\lambda_{e-ph})}\right)} \times f_1 \times f_2^* \qquad (4)$$

$$f_1 = \left(1 + \left(\frac{\lambda_{e-ph}}{2.46(1+3.8\mu^*)}\right)^{3/2}\right)^{1/3} \qquad (5)$$

$$f_2^* = 1 + (0.0241 - 0.0735\mu^*) \times \lambda_{e-ph}^2. \qquad (6)$$

where μ* is the Coulomb pseudopotential parameter (ranging from μ* = 0.10-0.16 [43-54]).

Eqs. 4-6 have a single solution in respect of $\lambda_{e-ph}$, for given $T_\theta$, $T_c$ and μ*. Due to μ* is varying within a range of 0.10-0.16 [43-54], in this work we used the mean value of μ* = 0.13 in all calculations. There is a need to clarify, that we define $T_c$ in Eqs. 4-6 based on the approach to be as closed as possible to the criterion [42]:

$$\frac{R(T)}{R(T_c^{onset})} \to 0.0 \qquad (7)$$

which is in given paper implemented as:

$$\frac{R(T)}{R(T_c^{onset})} = 0.03 \qquad (8)$$

It should stress that opposite definitions of $T_c$:

$$\frac{R(T)}{R(T_c^{onset})} \to 1.0 \qquad (9)$$

including widely used criteria of:

$$\frac{R(T)}{R(T_c^{onset})} = 0.5 \qquad (10)$$

$$\frac{R(T)}{R(T_c^{onset})} = 1.0 \qquad (11)$$

are not reliable, because Eqs. 10,11 cannot be used to distinguish the difference between the change in the resistance originated from any phase transition in the sample (for instance, the atomic ordering, the ferromagnetic ordering, structural phase transition, etc., see, for instance, Refs. 55-59) and the superconducting transition. The latter has one essential property which distinguishes the superconducting transition from all other phase transition, and this is zero resistance.

## 3. Results and Discussion

Equation 1 is in used to deduce the Debye temperature, $T_\theta$, in conductors from $R(T)$ data for more than six decades. For instance, we can mention the report by Hall *et al* [60] who fitted temperature dependent resistivity data for pure single crystal of yttrium and found that there is a very good agreement between $T_\theta$ deduced from the temperature dependent resistivity data and low-temperature heat capacity. More recently, Barker *et al* [61] reported on a very good agreement between $T_\theta$ values deduced from the resistivity data and from the specific heat data for noncentrosymmetric superconductor $Re_3Ta$. In regard of highly compressed superconductors, recently Matsumoto *et al* [62], reported a very good agreement between $\lambda_{e-ph}$ deduced from $R(T)$ data analysis in highly-compressed tin sulphide with the first-principles calculated $\lambda_{e-ph}$ reported by Gonzalez *et al* [63].

Despite we have reported results of application of Eq. 1 for the $R(T)$ data analysis for pure silver [64] and copper [64], as well as applications of Eqs. 1,3-6,8 for highly-compressed superconductors: black phosphorous [42], boron [42], oxygen [62], sulphur [39,62], lithium [62], GeAs [42], SiH$_4$ [42], BaH$_{12}$ [64], H$_3$S [42,65], D$_3$S [39,42], LaH$_x$ [39,42,65], LaD$_y$ [39,42], there is a need for additional confirmation, that our approach (Eqs. 1,3-6,8) is reliable research tool. To do that, in Sections 3.1-3.3 we presented analysis for elemental niobium (Section 3.1), intermetallic ReBe$_{22}$ alloy (Section 3.2), and elemental lanthanum (Section 3.3).

### 3.1. Elemental niobium

Webb [64] reported temperature dependence of the resistivity, $\rho(T)$, of ultra-pure elemental niobium which exhibited the residual resistance ratio of $\frac{R(T=298\ K)}{R(T=2\ K,\ B=0.7\ Tesla)} = 16,500$. In Figure 1 we showed reported $\frac{\rho(T)}{\rho(T=288\ K)}$ data together with the data fit to Eq. 1. Deduced $T_\theta = (252.6 \pm 0.8)\ K$ is in a middle of reported $T_\theta = (212 - 279)\ K$ values for pure niobium [67]. Based on reported $T_c$ = 9.25 K for this sample [64], we calculated $\lambda_{e-ph} = 0.85$ (for which we used Eqs. 4-6), which is in a proximity to $\lambda_{e-ph} = 0.82$ reported by McMillan [41] for niobium.

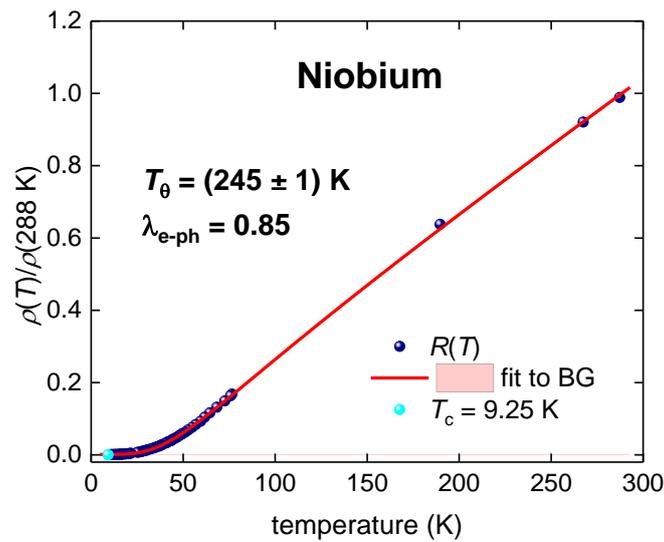

**Figure 1.** Resistivity data, $\rho(T)$, and fit to Eq. 1 for ultra-pure niobium. Raw $\rho(T)$ data reported by Webb [66]. Deduced $T_\theta = 245 \pm 1\ K$, $\lambda_{e-ph} = 0.85$; the fit quality is 0.99986. 95% confidence bands are shown by pink shaded area.

### 3.2. ReBe$_{22}$ alloy

Shang *et al* [68] reported temperature dependence of the resistivity, $\rho(T)$, of intermetallic ReBe$_{22}$ alloy, which exhibits 400-fold increase in T$_c$ in comparison with elemental beryllium. In Figure 2 we showed reported $\rho(T)$ data by Shang *et al* [68] together with the data fit to Eq. 3.

Shang *et al* [68] also performed low-temperature heat capacity studies of ReBe$_{22}$ alloy. From the results of those studies Shang *et al* [68] determined $\lambda_{e-ph} = 0.64(1)$, which is practically indistinguishable from our value of $\lambda_{e-ph} = 0.625$ deduced from $R(T)$ analysis in Figure 2. It also should be mentioned that in Ref. 67 we fitted the same $\rho(T)$ dataset to more advanced model [64,69-72]:

$$R(T) = R_0 + \theta(T_c^{onset} - T) \times \left( \frac{R(T_c^{onset})}{\left(I_0\left(F \times \left(1 - \frac{T}{T_c^{onset}}\right)^{3/2}\right)\right)^2} \right) + \theta(T - T_c^{onset}) \times \left( R(T_c^{onset}) + A \times \left( \left(\frac{T}{T_\theta}\right)^p \times \int_0^{\frac{T_\theta}{T}} \frac{x^p}{(e^x-1)(1-e^{-x})} dx - \left(\frac{T_c^{onset}}{T_\theta}\right)^p \times \int_0^{\frac{T_\theta}{T_c^{onset}}} \frac{x^p}{(e^x-1)(1-e^{-x})} dx \right) \right) \quad (12)$$

where $p$ is free-fitting parameter and $T_\omega$ is integrated characteristic temperature of the charge carrier scattering mechanism. It should be mentioned that $R(T)$ curves for MgB$_2$ [70], ferromagnetic metals and materials [69,71], highly-compressed iron [69], and some other materials [64,69,72], were analysed by using advanced Bloch-Grüneisen equation (Eq. 12). However, in this work, we analysed data in the assumption of pure electron-phonon scattering mechanism and, thus, we fixed $p$ value to $p = 5$ and $T_\omega \rightarrow T_\theta$ (more details can be found elsewhere [64,69-72]).

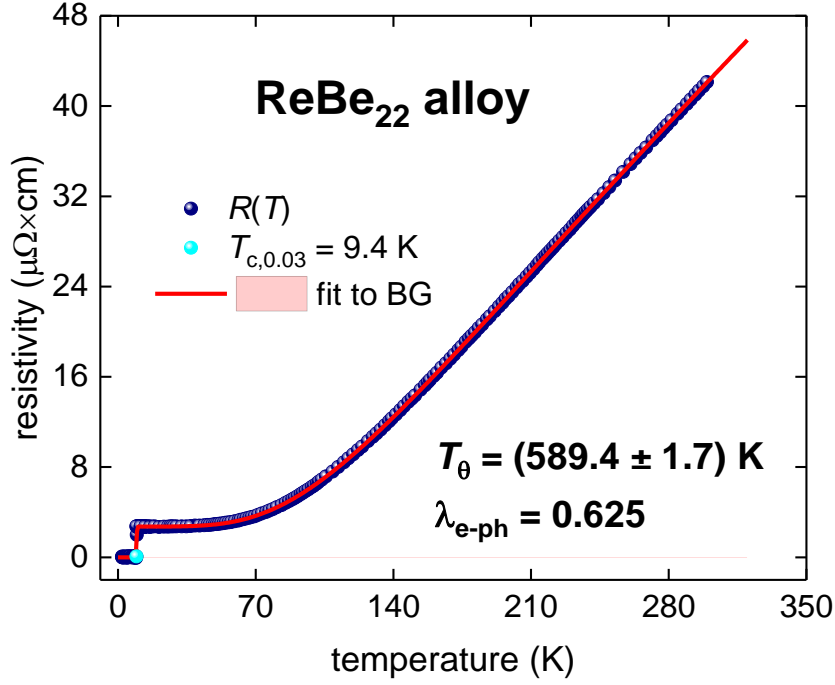

**Figure 2.** Resistivity data, $\rho(T)$, and fit to Eq. 1 for ReBe$_{22}$ alloy. Raw $\rho(T)$ data reported by Shang *et al* [68]. Deduced $T_\theta = 589.4 \pm 1.7\,K$, $\lambda_{e-ph} = 0.625$; the fit quality is 0.9998. 95% confidence bands are shown by pink shaded area.

### 3.3. Elemental lanthanum

Alstad *et al* [73] reported temperature dependence of the resistivity, $\rho(T)$, of lanthanum (which is shown in Figure 3). To analyse $\rho(T)$ data for this element, we applied so-called saturated resistivity model proposed by Fisk and Webb [74], and Wiesmann *et al* [75], to fit $\rho(T)$ data for A-15 alloys (including, Nb$_3$Sn [74,75]). The expression for the model is [75]:

$$\rho(T) = \cfrac{1}{\cfrac{1}{\rho_{sat}} + \cfrac{1}{\rho_0 + A \times \left(\cfrac{T}{T_\theta}\right)^5 \times \int_0^{\frac{T_\theta}{T}} \cfrac{x^5}{(e^x-1)(1-e^{-x})} dx}} \qquad (13)$$

where $\rho_{sat}$ is free-fitting parameter. The fit of $\rho(T)$ data to Eq. 13 is shown in Figure 3.

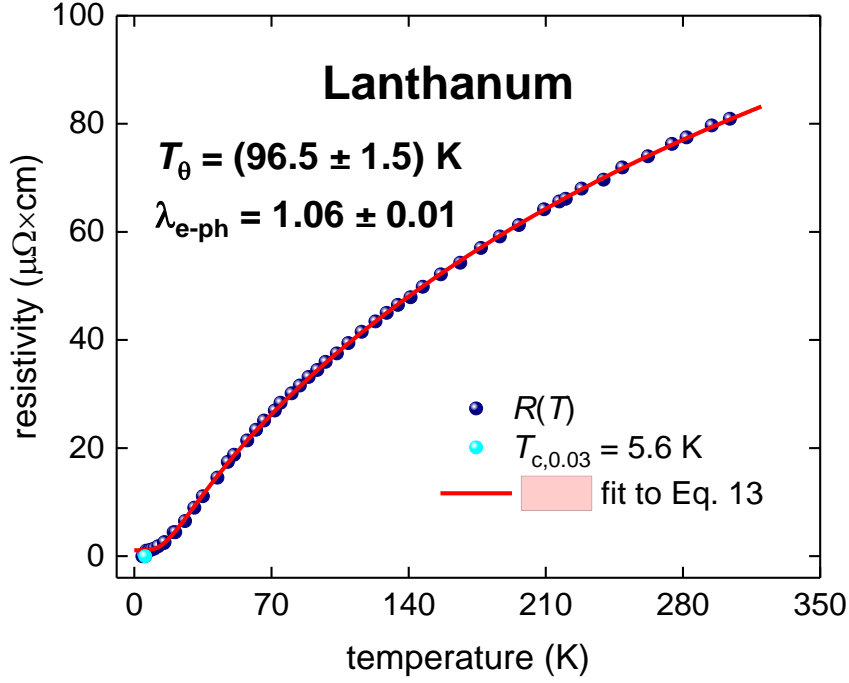

**Figure 3.** Resistivity data, $\rho(T)$, and fit to Eq. 13 for lanthanum. Raw $\rho(T)$ data reported by Alstad *et al* [73]. Deduced $T_\theta = 56.5 \pm 1.5\ K$, $\lambda_{e-ph} = 1.06 \pm 0.01$ (for $\mu^* = 0.13$); and $\lambda_{e-ph} = 0.97 \pm 0.01$ (for $\mu^* = 0.10$); $\rho_{sat} = 180 \pm 3\ \Omega$; fit quality is 0.99995. 95% confidence bands are shown by pink shaded area.

Bağcı *et al* [76] reviewed results of first-principles calculations for lanthanum and showed that $\lambda_{e-ph}$ is varied in the range of $\lambda_{e-ph} = 0.83 - 0.97$. Our value of $\lambda_{e-ph} = 1.06 \pm 0.01$ was calculated in the assumption of $\mu^* = 0.13$, however, if one can use $\mu^* = 0.10$, then $\lambda_{e-ph} = 0.97 \pm 0.01$, which is exact value calculated by Bağcı *et al* [76] for lanthanum. It should be noted, that recently Chen *et al* [77] performed experimental and first-principles studies of superconducting state of the lanthanum up to megabar pressure. However, analysis of experimental *R*(*T*) data vs applied pressure for lanthanum is beyond the topic of this paper.

Overall, results of $\rho(T)$ analysis for niobium, ReBe$_{22}$, and lanthanum confirm (in addition to previously published results [42,65]) that Eqs. 1,3-6,8 are reliable research tool. In following Sections, we analysed temperature dependent resistance in polyhydrides by utilizing the same approach.

### 3.4. Th$_4$H$_{15}$ at ambient pressure

Satterthwaite and Peterson synthesised massive samples of $I\bar{4}3d$-Th$_4$H$_{15}$ phases and studies the effect of vacuum anneal on the superconducting transition of this phase [12]. In Fig. 4 we show resistance curves and data fit to Eqs. 1,3 for "as reacted" (Fig. 4(a)), vacuum anneal at $T$ = 150 C (Fig. 4(b)), and vacuum anneal at $T$ = 295 C (Fig. 4(c)) Th$_4$H$_{15}$ samples.

It should be noted that transition temperature for "as reacted" sample was not reported, however Satterthwaite and Peterson [12] found that the vacuum anneal causes the decrease of the hydrogen concentration in of $I\bar{4}3d$-Th$_4$H$_{15}$ phase. Returning to our discussion on the $T_c$ criterion definition (Eqs. 7-11 and Ref. [40]), it should be noted that hydrogen stoichiometry does not make any significant impact on $T_c^{onset}$ (Fig. 4(b,c)). However, it significantly changes $T_{c,0.03}$ and, thus, this is another evidence that $T_c^{onset}$ is not reliable value for the $T_c$ definition.

Deduced electron-phonon coupling constant, $\lambda_{e-ph}$, is in a reasonable agreement with the value of $\lambda_{e-ph} = 0.84$ reported by Miller *et al* [78], who used low-temperature heat capacity measurements to deduce the constant. In Fig. 5 we summarized reported [78] and deduced, by the *R*(*T*) analysis herein, $\lambda_{e-ph}$ values for $I\bar{4}3d$-Th$_4$H$_{15}$ phase.

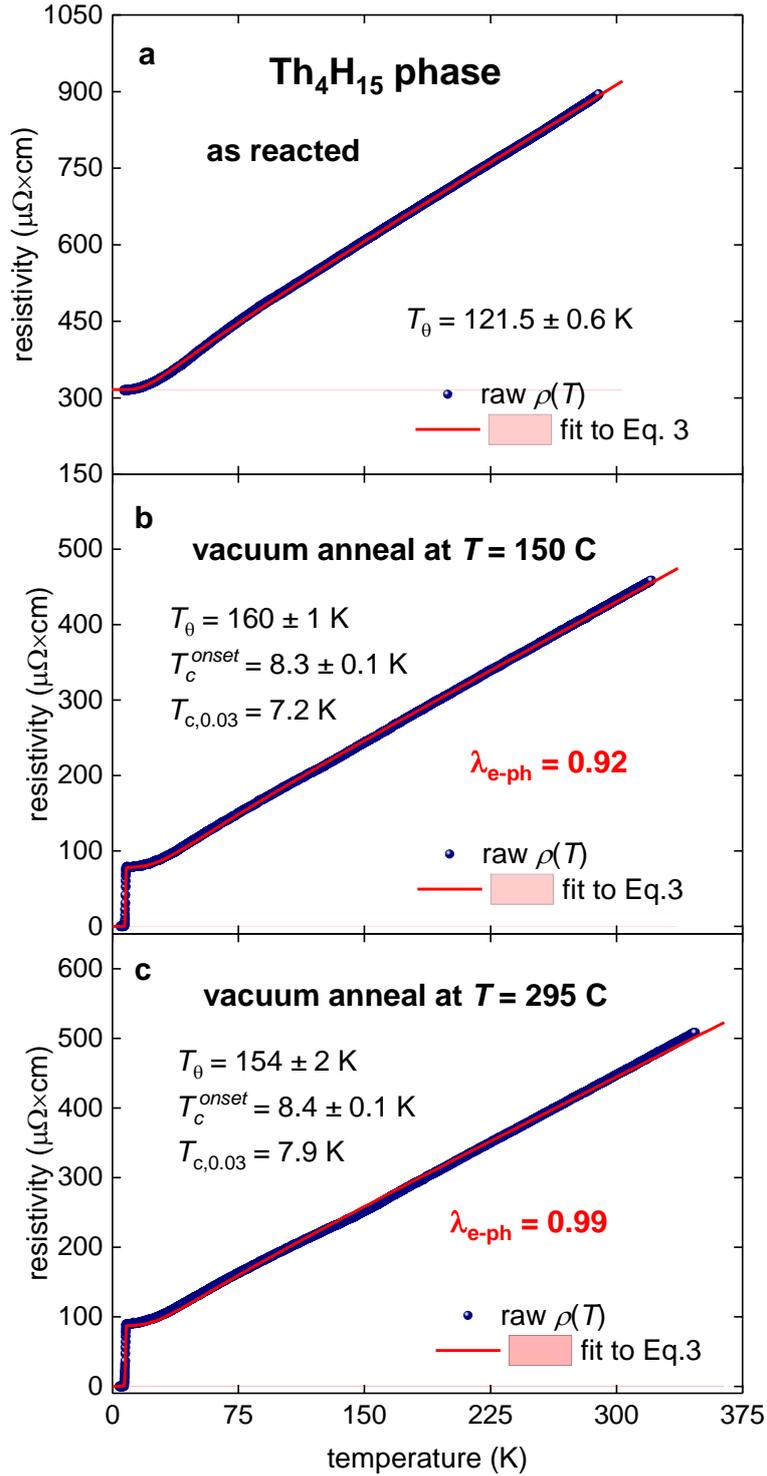

**Figure 4.** Resistivity data, $\rho(T)$, and fit to Eq. 1 (for panel a) and to Eq. 3 (for panels b and c) for massive $I\bar{4}3d$-Th$_4$H$_{15}$ samples. Raw $\rho(T)$ data reported by Satterthwaite and Peterson [12]. **a** – "as reacted" sample; deduced $T_\theta = 121.5 \pm 0.1\ K$; the fit quality is 0.99991. **b** – vacuum annealed sample at $T = 190$ C, deduced $T_c^{onset} = 8.3 \pm 0.1\ K$, $T_\theta = 160 \pm 1\ K$; $T_{c,0.03} = 7.2\ K$, $\lambda_{e-ph} = 0.92$; the fit quality is 0.9998. **c** – vacuum annealed sample at $T = 295$ C, deduced $T_c^{onset} = 8.4 \pm 0.1\ K$, $T_\theta = 154 \pm 2\ K$; $T_{c,0.03} = 7.9\ K$, $\lambda_{e-ph} = 0.99$; the fit quality is 0.9990. 95% confidence bands are shown by pink shaded areas in all panels.

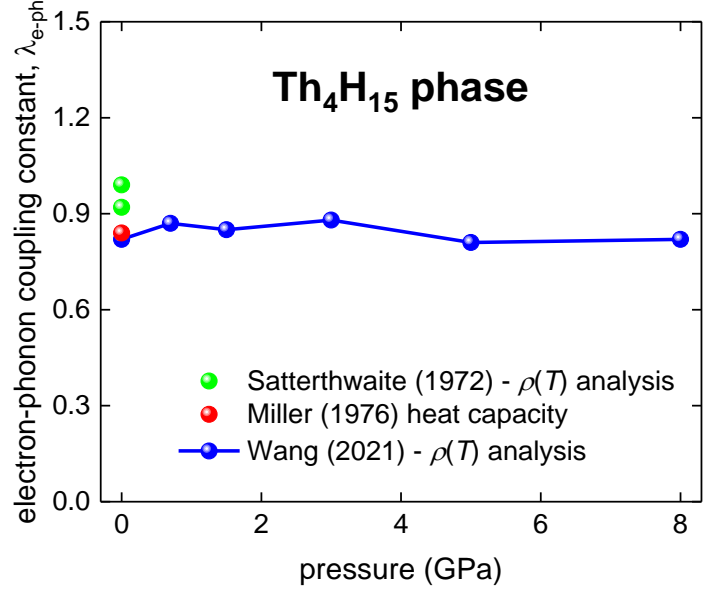

**Figure 5.** The electron-phonon coupling constant, $\lambda_{e-ph}$, for $I\bar{4}3d$-$Th_4H_{15}$. Green balls are deduced values from the analysis of $\rho(T)$ data presented in Section 3.1 herein (raw data is from Satterthwaite and Peterson [12]). Red ball is reported value by Miller *et al* [78]. Blue balls are deduced values from the analysis of $\rho(T)$ data presented in Section 3.2 herein (raw data reported by Wang *et al* [79]).

### 3.5. $Th_4H_{15}$ at high pressure

Recently, Wang *et al* [79] synthesized $I\bar{4}3d$-$Th_4H_{15}$ phase via hydrogenating thorium metal at $P = 5$ GPa and $T = 800$ °C by using the ammonia borane as the hydrogen source for the reaction. This research group [79] has measured the temperature-dependent resistivity $\rho(T)$ under different hydrostatic pressures from 0 to 8 GPa by using a cubic anvil cell apparatus, and performed measurements of magnetic susceptibility and specific heat at ambient pressure. Before analysing the temperature-dependent resistivity data, we first point out an unintentional mistake in Ref. [79] that leads to incorrect estimations of the London penetration depth λ and the Ginzburg-Landau (GL) parameter $\kappa$ at 2 K for $I\bar{4}3d$-$Th_4H_{15}$ phase.

Based on the magnetoresistance $\rho(T, B)$ data and the criterion of 50% of normal-state resistance for $T_c$, Wang *et al* [79] deduced the $B_{c2}(T)$ data showed in their Fig. 3(c) and fit to the upper critical field model proposed by Jones *et al* [80]:

$$B_{c2}(T) = \frac{\phi_0}{2\pi\xi^2(0)} \times \left(\frac{1-\left(\frac{T}{T_c}\right)^2}{1+\left(\frac{T}{T_c}\right)^2}\right) \tag{14}$$

where, $\phi_0 = 2.067 \cdot 10^{-15}$ Wb is the magnetic quantum flux, yielding the zero-temperature coherence length ξ(0 K) = 8.36 nm. Then, based on the magnetization data at 2 K, Wang *et al* [79] had estimated the lower critical field $B_{c1}(T = 2$ K$) = 20$ mT in their Fig. 2(b), and deduced the London penetration depth, λ(*T* = 2 K) = 8.43 nm by employing the widely used expression [81,82]:

$$B_{c1}(T) = \frac{\phi_0}{4\pi\lambda^2(T)} \times \left(ln\left(\frac{\lambda(T)}{\xi(T)}\right)\right) \tag{15}$$

and a fixed value of $\xi(T = 0K) = 8.36 \, nm$ obtained above.

Since the London penetration depth λ was estimated based on the $B_{c1}$ data at 2 K, Wang et al. [79] should use the coherence length at 2 K, $\xi(T = 2 \, K) = 9.7 \, nm$ based on the extrapolated $B_{c2}(T = 2 \, K) = 3.5 \, T$ (Fig. 3(c) in Ref. 79) according to the GL equation:

$$\xi(T) = \sqrt{\frac{\phi_0}{2\pi B_{c2}(T)}} \tag{16}$$

Even after making this correction, however, the obtained λ(*T* = 2 K) = 9.7 nm is too small to be physically meaningful. After communicating with the authors of Ref. [79,83], they pointed out that Eq. 15 has two solutions with respect of $B_{c1}(T)$ as schematically showed in Figure 6, and they have taken the incorrect smaller solution. However, a unique solution has been generally expected for Eq. 15 [81,84]. As shown in Fig. 6, the same issue also exists even if a more accurate expression for the lower critical field is used [85-87]:

$$B_{c1}(T) = \frac{\phi_0}{4\pi\lambda^2(T)} \times \left(ln\left(\frac{\lambda(T)}{\xi(T)}\right) + 0.5\right) \tag{17}$$

The same problem was found for very complicated $B_{c1}(T)$ formula proposed by Brant [84] (Fig. 6):

$$B_{c1}(T) = \frac{\phi_0}{4\pi\lambda^2(T)} \times \left(\ln\left(\frac{\lambda(T)}{\xi(T)}\right) + \alpha\left(\frac{\lambda(T)}{\xi(T)}\right)\right) \tag{18}$$

where

$$\alpha(\kappa) = \alpha_\infty + e^{\left(-c_0 - c_1 \times ln\left(\frac{\lambda(T)}{\xi(T)}\right) - c_2 \times \left(ln\left(\frac{\lambda(T)}{\xi(T)}\right)\right)^2\right)} \pm \varepsilon \qquad (19)$$

where $\alpha_\infty = 0.49693$, $c_0 = 0.41477$, $c_1 = 0.775$, $c_2 = 0.1303$, and $\varepsilon \leq 0.00076$ [84].

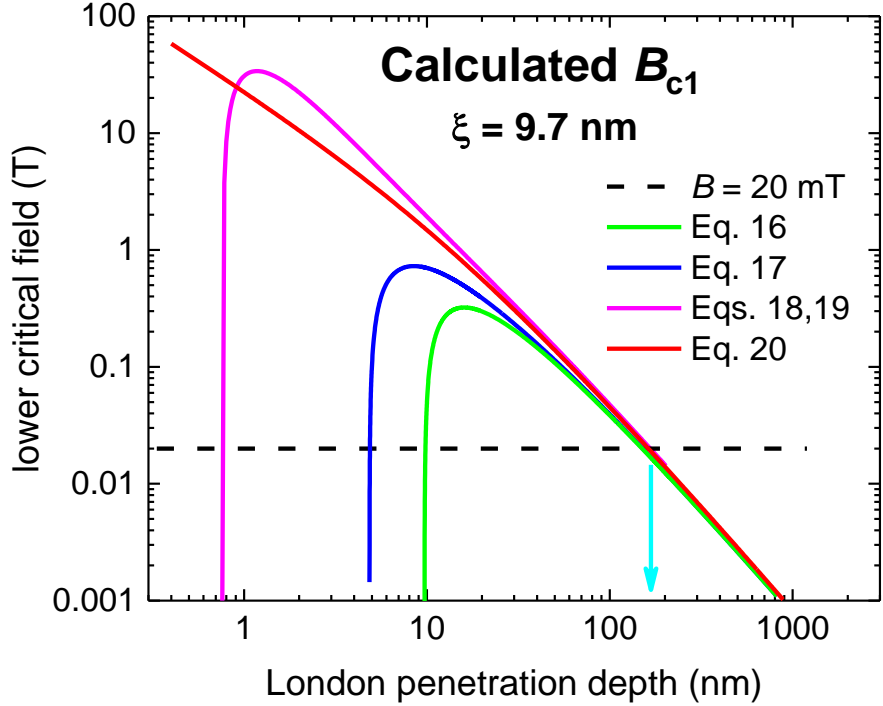

**Figure 6.** Calculated lower critical field, $B_{c1}$, at fixed superconducting coherence length, $\xi$ = 9.7 nm, by Eqs. 15,17-20. Unique solution in respect of London penetration depth calculated by Eq. 20 is shown by blue arrow. Courtesy to Prof. Jinguang Cheng [83] (Institute of Physics, Chinese Academy of Sciences) for the idea of graphical solution of Eqs. 15,17-20.

In Ref. 88, the author has proposed a universal equation for the Meissner-Ochsenfeld (MO) field, which is the thermodynamic field, $B_c$, for Type-I superconductors, and the lower critical field, $B_{c1}$, in Type-II superconductors:

$$B_{MO}(T) = \frac{\phi_0}{4\pi\lambda^2(T)} \times \left(1 + \sqrt{2} \times ln\left(\frac{\lambda(T)}{\xi(T)}\right)\right). \qquad (20)$$

It is interesting to note that this equation has a unique solution for $\lambda(T = 2\,K) = 163\,nm$, as also illustrated in Figure 6. By using the correct value of $\lambda(T = 2\,K)$, the GL parameter $\kappa(T = 2\,K) = \frac{\lambda(T=2\,K)}{\xi(T=2\,K)} = 17$ is obtained. Detailed discussion of this problem can be found elsewhere [88,89]. More careful examination of Fig. 2(b) in Ref. 76 gives $B_{c1}(T = 2\,K) = 11$ mT, from which $\lambda(T = 2\ K) = 230\ nm$ and $\kappa(T = 2\ K) = 24$ can be obtained. Thus, there is an additional advantage to use universal Eq. 18 for calculations as the thermodynamic field in Type-I superconductors, as the lower critical field in Type-II superconductors.

In Fig. 7 we fitted $\rho(T)$ datasets reported by Wang *et al* [79] to Eq. 3. Deduced electron-phonon coupling constant, $\lambda_{e-ph}$, for $I\bar{4}3d$-$Th_4H_{15}$ vs applied pressure is shown in Fig. 5, where it can be seen that $\lambda_{e-ph}$ remains nearly constant within full pressure range and all deduced $\lambda_{e-ph}$ values are in excellent agreement with $\lambda_{e-ph}$ reported by Miller *et al* [78].

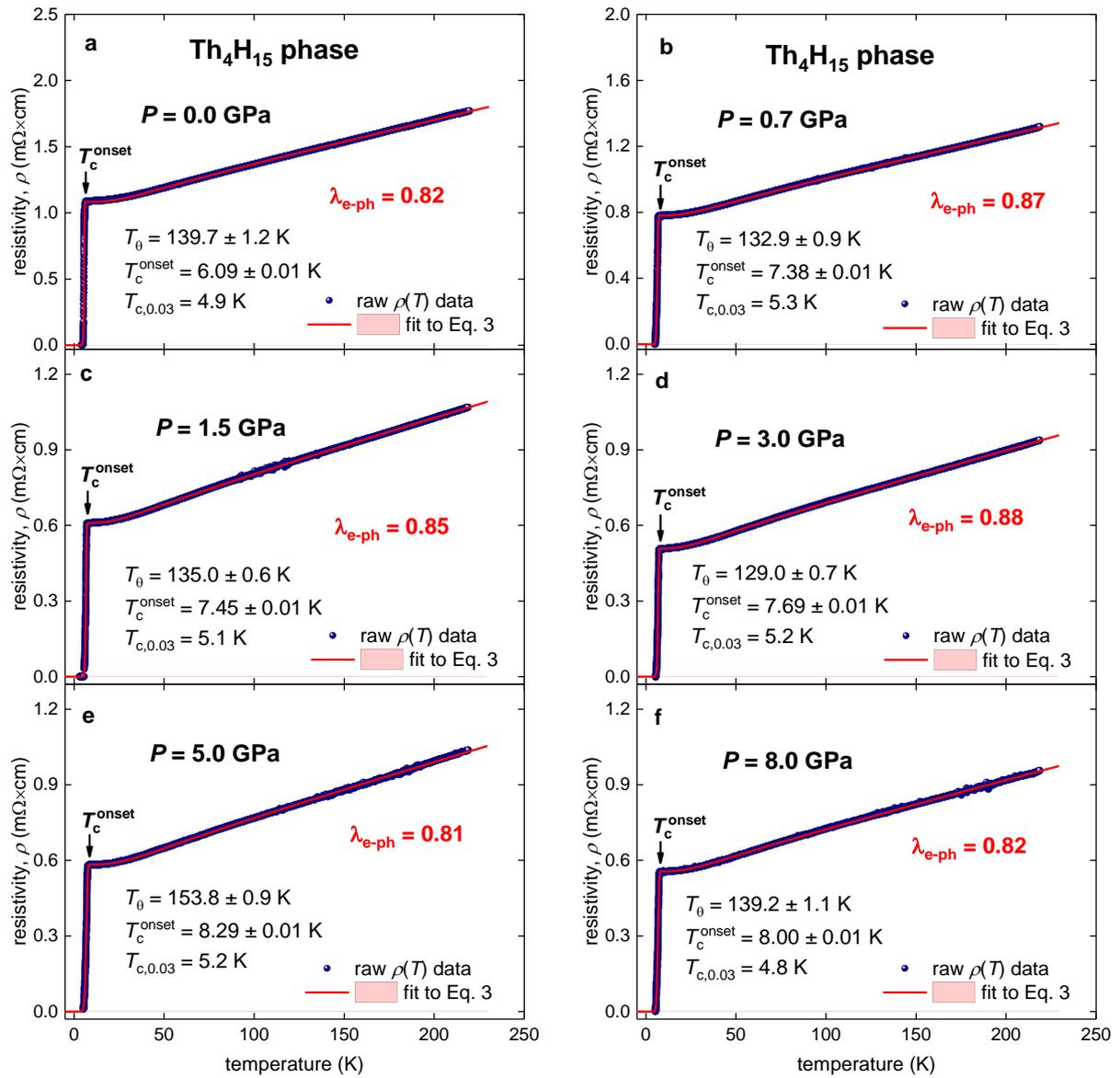

**Figure 7.** Resistivity data, $\rho(T)$, and fit to Eq. 3 for $I\bar{4}3d$-$Th_4H_{15}$ samples reported by Wang *et al* [79]. (a) ambient pressure, deduced $\lambda_{e-ph} = 0.82$, $T_\theta = 139.7 \pm 1.2\ K$, fit quality is 0.9992; (b) $P = 0.7$ GPa, deduced $\lambda_{e-ph} = 0.87$, $T_\theta = 132.9 \pm 0.9\ K$, fit quality is 0.9996;

(c) $P = 1.5$ GPa, deduced $\lambda_{e-ph} = 0.85$, $T_\theta = 135.0 \pm 0.6\ K$, fit quality is 0.9997. (d) $P = 3.0$ GPa, deduced $\lambda_{e-ph} = 0.88$, $T_\theta = 129.0 \pm 0.7\ K$, fit quality is 0.9997; (e) $P = 5.0$ GPa, deduced $\lambda_{e-ph} = 0.81$, $T_\theta = 153.8 \pm 0.9\ K$, fit quality is 0.9995. (f) $P = 8.0$ GPa, deduced $\lambda_{e-ph} = 0.82$, $T_\theta = 139.2 \pm 1.1\ K$, fit quality is 0.9994. 95% confidence bands are shown by pink shadow areas.

It should be also stressed that $\lambda_{e-ph}$ values deduced from experiment (Fig. 7) are in notable disagreement with $\lambda_{e-ph} = 0.38$ reported by Shein *et al* [90] who computed the latter through an estimated value of the Sommerfeld coefficient, γ, assuming the free-electron model. Thus, there is a request for first-principles calculations for the first superconducting superhydride $I\bar{4}3d$-Th$_4$H$_{15}$ discovered five decades ago.

### 3.6. Highly-compressed ThH₉

Semenok *et al* [13] reported on the discovery two high-temperature superconducting thorium superhydride phases, $P6_3/mmc$-ThH₉ ($T_c \cong 145\ K$) and $Fm\bar{3}m$-ThH₁₀ ($T_c \cong 160\ K$). In their Figure 5(b) Semenok *et al* [13] reported $R(T)$ dataset for $P6_3/mmc$-ThH₉ phase compressed at $P = 170$ GPa, which we fitted to Eq. 1 in Fig. 8. The fit converged at $T_\theta = 1453 \pm 13\ K$ and deduced $\lambda_{e-ph}(170\ GPa) = 1.46 \pm 0.01$.

Despite Semenok *et al* [13] did not report the value for $\lambda_{e-ph}(170\ GPa)$ for $P6_3/mmc$-ThH₉ phase, they computed $\lambda_{e-ph}(100\ GPa) = 2.15$ and $\lambda_{e-ph}(150\ GPa) = 1.73$ [13]. Linear extrapolation, based on these two values, gives for $\lambda_{e-ph}(170\ GPa) = 1.57$, which is in a good proximity to our value of $\lambda_{e-ph}(170\ GPa) = 1.46 \pm 0.01$. Thus, we can conclude that first-principles calculations reported by Semenok *et al* [13] can be characterized as accurate theoretical approach to study highly-compressed thorium superhydrides.

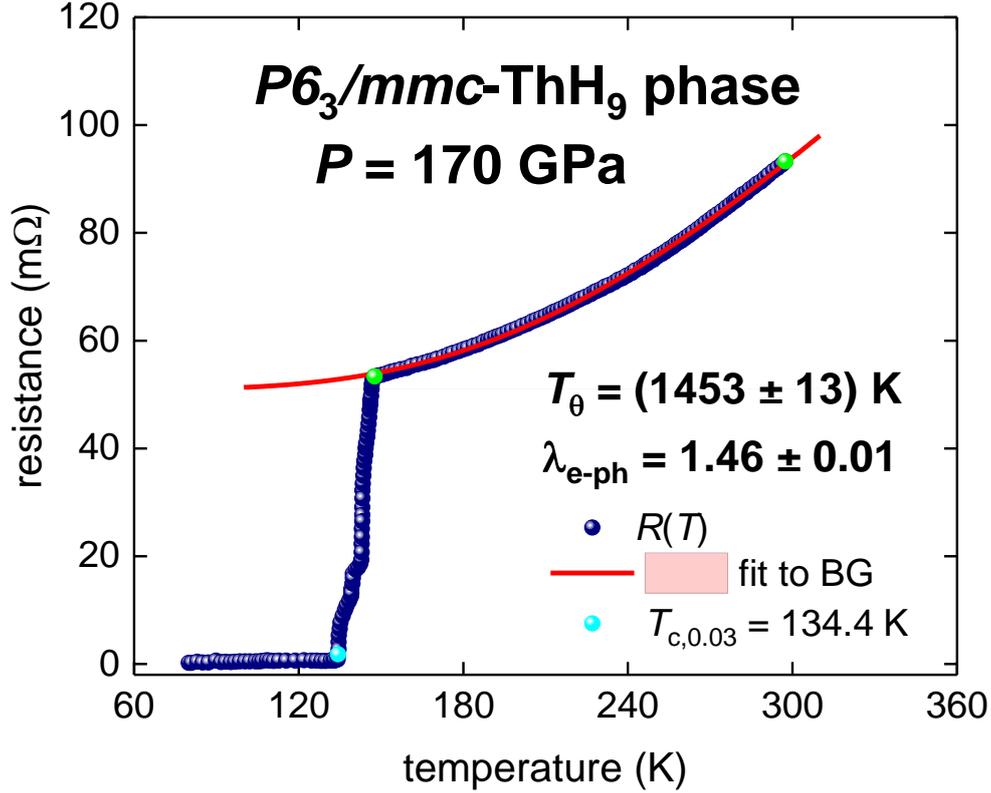

**Figure 8.** Resistance data, $R(T)$, and data fits to Eq. 1 for highly-compressed $P6_3/mmc$-ThH$_9$ ($P$ = 170 GPa) (raw data is from Ref. 13). Green balls indicate the bounds for which $R(T)$ data was used for the fit to Eq. 1. Deduced $T_\theta = 1453 \pm 13\ K$, $T_{c,0.03} = 134.4\ K$, $\lambda_{e-ph} = 1.46 \pm 0.01$, fit quality is 0.9995. 95% confidence bands are shown by pink shadow areas.

### 3.7. Highly-compressed ThH$_{10}$

Semenok *et al* [13] also reported $R(T)$ dataset for $Fm\overline{3}m$-ThH$_{10}$ compressed at $P = 170$ GPa in their Figure S31(b). Raw $R(T)$ data displayed in Figure 9 was provided by D. V. Semenok (Skolkovo Institute of Science and Technology). Two $R(T)$ datasets were measured by different contacts arrangements for the same sample. These configurations designated as Channel 2 and Channel 4 in Fig. 9, where data fits to Eq. 1 are shown.

Both deduced $\lambda_{e-ph}(170\ GPa) = 1.71 \pm 0.03$ (for Channel 2 (Fig. 9,a) and $\lambda_{e-ph}(170\ GPa) = 1.68 \pm 0.02$ (for Channel 4 (Fig. 9,b) are in remarkable agreement with computed by first-principles calculation value of $\lambda_{e-ph}(174\ GPa) = 1.75$ for $Fm\overline{3}m$-ThH$_{10}$ phase by Semenok *et al* [13]. This result jointly confirms as calculations by Semenok *et al* [13], as our approach based on Eqs. 1,4-6,8 (which was proposed in Ref. [42]).

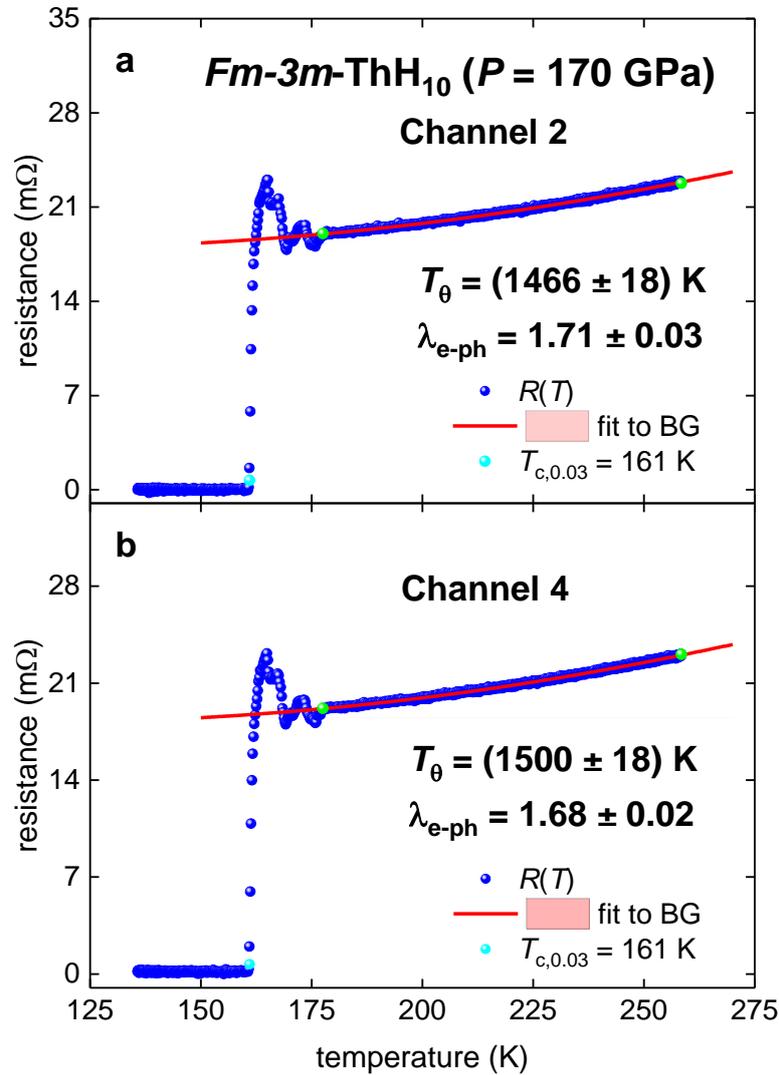

**Figure 9.** Resistance data, *R(T)*, and data fits to Eq. 1 for highly-compressed $Fm\bar{3}m$-ThH$_{10}$ phase (*P* = 170 GPa) (raw data reported by Semenok *et al* [13]). Green balls indicate the bounds for which *R(T)* data was used for the fit to Eq. 1. (a) Deduced $T_\theta = 1466 \pm 18\,K$, $T_{c,0.03} = 161\,K$, $\lambda_{e-ph} = 1.71 \pm 0.03$. (b) Deduced $T_\theta = 1500 \pm 18\,K$, $T_{c,0.03} = 161\,K$, $\lambda_{e-ph} = 1.68 \pm 0.02$. Fits goodness for both panels > 0.997. 95% confidence bands are shown by pink shadow areas.

### 3.8. Highly-compressed YD$_6$

Troyan *et al* [20] reported on the discovery of near-room temperature superconductivity (NRTS) in highly-compressed polyhydrides/polydeuterides of yttrium. Nearly simultaneously,

Kong et al [21] also reported on the observation of NRTS in highly-compressed yttrium polyhydrides/polydeuterides. Here in Figure 10 we showed the fit of $R(T)$ data for *Im-3m*-YD$_6$ phase ($P$ = 172 GPa) to Eq. 1. Raw $R(T)$ data for Figure 10 was provided by A. G. Kvashnin (Skolkovo Institute of Science and Technology).

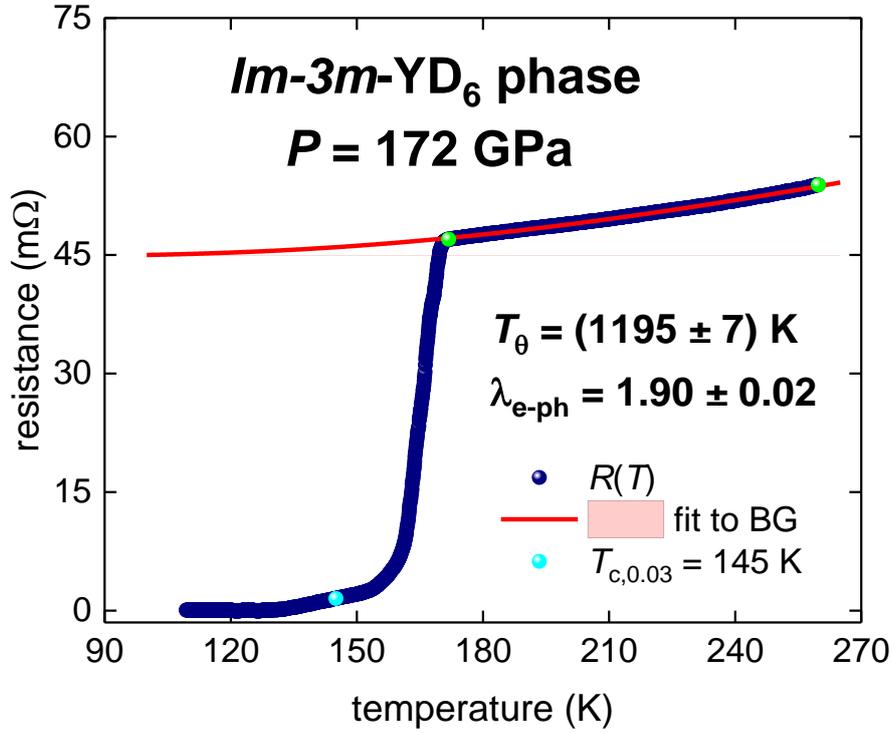

**Figure 10.** Resistance data, $R(T)$, and data fits to Eq. 1 for highly-compressed *Im-3m*-YD$_6$ phase ($P$ = 172 GPa) (raw data reported by Troyan et al [20]). Green balls indicate the bounds for which $R(T)$ data was used for the fit to Eq. 1. Deduced $T_\theta = 1195 \pm 7\ K$, $T_{c,0.03} = 145\ K$, $\lambda_{e-ph} = 1.90 \pm 0.02$. Goodness of fit is 0.9991. 95% confidence bands are shown by pink shadow areas.

It should be noted that Troyan et al [20] reported the first-principles calculated $\lambda_{e-ph}(165\ GPa) = 1.80$ for *Im-3m*-YD$_6$. This value was computed by including the anharmonic term in first-principles calculations (see, Table S8 in Supplementary Information of Ref. 20). This computed value is in excellent agreement with the value of $\lambda_{e-ph}(172\ GPa) = 1.90 \pm 0.02$ which we deduced from $R(T)$ dataset by our approach (Eqs. 4-6,8).

### 3.9. Metallized hydrogen phase III

Eremets *et al* [36] reported first $R(T)$ curve for metallic hydrogen phase III (at P = 402 GPa) measured by four-probe technique. We analysed this $R(T)$ curve in Fig. 11. Deduced $T_\theta$(402 GPa) = 727 ± 6 K. Superconducting transition was not observed at pressure in the range of $P$ = 300-402 GPa [36], while samples were cooled down to $T$ = 80 K (these measurements were performed by two-probe technique). This implies that hydrogen phase III exhibits the electron-phonon coupling constant of $\lambda_{e-ph}(402\ GPa) \leq 1.7$ (we used $\mu^* = 0.13$ to calculate this value). This $\lambda_{e-ph}(402\ GPa)$ upper bound is well matched the range of $\lambda_{e-ph}$ recently computed by Dogan *et al* [93], who reported that three potentially possible phases of metallic hydrogen should exhibit $\lambda_{e-ph}(400\ GPa) = 0.67 - 1.73$.

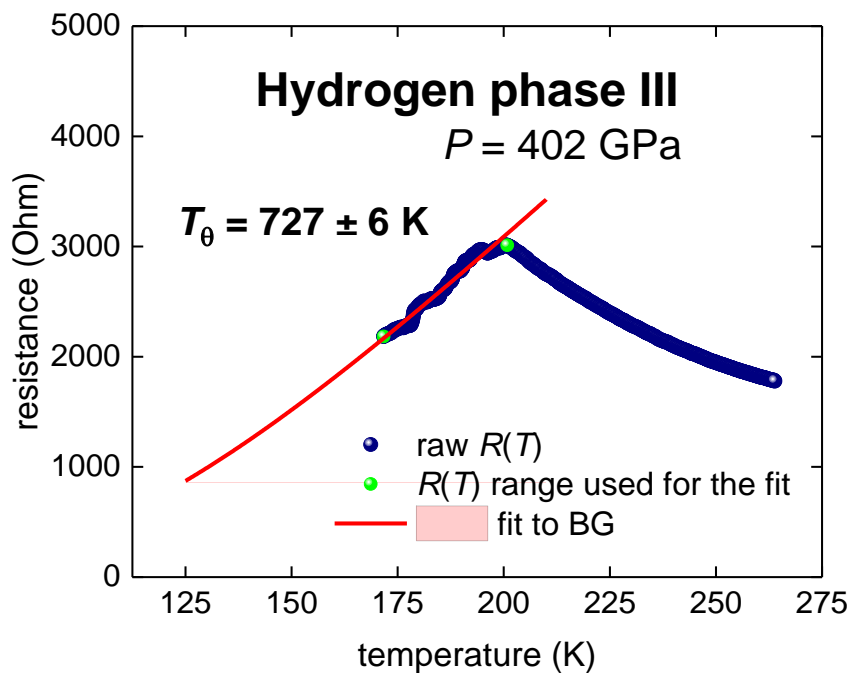

**Figure 11.** Resistance data, $R(T)$, and data fits to Eq. 1 for highly-compressed hydrogen phase III ($P$ = 402 GPa) (raw data reported by Eremets *et al* [34]). Green balls indicate the bounds at which $R(T)$ data was used for the fit to Eq. 1. Deduced $T_\theta = 727 \pm 6\ K$, $\lambda_{e-ph} \leq 1.7$. Goodness of fit is 0.974. 95% confidence bands are shown by pink shadow areas.

## 4. Summary

In Table I we summarised all our results of application of Bloch-Grüneisen equation [37,38] for the condition of $p = 5$ (Eqs. 1,4-6,8) [42,64,65,72]. It can be seen (Table I) that there is a good agreement between $\lambda_{e-ph}$ deduced by Eqs. 1,4-6,8 and $\lambda_{e-ph}$ computed by first-principles calculations or deduced from experimental data different from the temperature dependent resistance analysis. In Table I we referred the first-principles calculated $\lambda_{e-ph}$ values which are the most close prediction/confirmation for experimentally observed results.

**Table I.** Deduced $T_c$, $T_\theta$, $\lambda_{e-ph}$ for several metallic materials from $R(T)$ data analysis for which Eqs. 1,4-6,8 were used. Independently reported $\lambda_{e-ph}$ computed by first-principles calculations or deduced from experiment are also included.

| element/compound ($R(T)$ data source) | pressure (GPa) | $T_{c,0.03}$ (K) | used $\mu^*$ | deduced $T_\theta$ (K) | deduced $\lambda_{e-ph}$ | independently reported $\lambda_{e-ph}$ |
|---|---|---|---|---|---|---|
| niobium [66] | ambient | 9.25 | 0.13 | 245±1 [here] | 0.85 [here] | 0.82–1.26 [92] |
| ReBe$_{22}$ [68] | ambient | 9.5 | 0.13 | 591±8 [here] | 0.63 [here] | 0.64 [68] |
| lanthanum [73] | ambient | 5.6 | 0.10 | 96.5±1.5 [here] | 0.97±0.01 [here] | 0.97 [76] |
| $I\bar{4}3d$ Th$_4$H$_{15}$ [12] | ambient | 0-7.9 | 0.13 | 120-160 [here] | 0.92-0.99 [here] | 0.84 [78] |
| $I\bar{4}3d$ Th$_4$H$_{15}$ [79] | 0-8 | 4.8-5.4 | 0.13 | 129-154 [here] | 0.81-0.88 [here] | 0.84 [78] |
| $P6_3/mmc$-ThH$_9$ [13] | 170 | 134 | 0.13 | 1453±13 [here] | 1.46±0.01 [here] | 1.57 (linear extrapolated value to 170 GPa) [13] |
| $Fm\bar{3}m$-ThH$_{10}$ (channel 2) [13] | 170 | 161 | 0.13 | 1466±18 [here] | 1.71±0.03 [here] | 1.75 [13] (174 GPa) |
| $Fm\bar{3}m$-ThH$_{10}$ (channel 4) [13] | 170 | 161 | 0.13 | 1500±18 [here] | 1.68±0.02 [here] | 1.75 [13] (174 GPa) |
| $Im$-$3m$-YD$_6$ [20] | 172 | 145 | 0.13 | 1195±7 [here] | 1.90±0.02 [here] | 1.80 [20] (165 GPa) |
| Hydrogen phase III [34] | 402 | none | 0.13 | 727±6 [here] | <1.7 [here] | 0.67-1.73 [91] (400 GPa) |
| Sulphur [93] | 110 | 8.7 | 0.13 | 221±2 [65] | 0.87 ($P$ = 110 GPa) [65] | 0.65 ($P$ = 500 GPa) [94] |
| black phosphorus [95] | 15 | 5.3 | 0.17 [96] | 563±6 [40] | 0.628±0.005 [42] | 0.627-0.673 [96] |
| boron [97] | 240 | 7.4 | 0.12 | 314±2 [42] | 0.68 [42] | 0.39 [98] (215-279 GPa) |
| | | | $(R(T)/R_{norm}=0.67)$ | | | |
| GeAs [99] | 20.6 | 3.95 | 0.10 [99] | 405±4 [42] | 0.493 [42] | 0.52-0.65 [99] 20 GPa |
| SiH$_4$ [100] | 192 | 7.7 | 0.12 [101] $(R(T)/R_{norm}=0.64)$ | 353±3 [40] | 0.67 [42] | 0.8 [101] |
| $R3m$-H$_3$S [102] | 133 | 164.8 | 0.13 [39] | 1153±44 [37] | 2.3±0.1 [39] | 2.07 [103] 130 GPa |
| $Im\bar{3}m$-D$_3$S [102] | 190 | 127.2 | 0.13 [39] | 1070±6 [37] | 1.85 [39] | 1.87 [47] 200 GPa |
| $Im\bar{3}m$-H$_3$S [104] (07/2018) | 155 | 192.5 | 0.10 [42] | 1426±1 [42] | 1.93 [42] | 1.84 [47] 200 GPa |
| LaH$_x$ (x > 3), Sample 11 [18] | 150 | 64.4 | 0.10 [39] | 599±1 [39] | 1.68 [39] | 0.845-2.18 [105,106] |
| $Fm\bar{3}m$ LaH$_{10}$, Sample 3 [18] | 150 | 240 | 0.10 [42] | 1310±9 [42] | 2.77±0.02 [42] | 2.76 [43] 163 GPa |
| pseudocubic BaH$_{12}$ [30] | 100 | 5.1 | 0.13 | 348±2 [64] | 0.61 [here] | 0.95 [30] 140 GPa |

Full $\lambda_{e-ph}$ dataset from Table I is fitted to the linear function in Figure 12, where it can be seen a good agreement between reported and deduced (by the approach described herein and in Refs. 39,42) $\lambda_{e-ph}$ values. In Figure 12 we also depicted two cases for which the difference is large, as we discussed above.

In conclusion, in this paper we analysed temperature dependent resistance data for known to date superconducting polyhydrides of thorium which were reported over five decades, since 1971. We found a very good agreement between the electron-phonon coupling constant, $\lambda_{e-ph}$, for $I\bar{4}3d$-Th$_4$H$_{15}$ phase deduced from $R(T)$ and from low-temperature heat capacity measurements. In addition, we showed that $\lambda_{e-ph}$ in $I\bar{4}3d$-Th$_4$H$_{15}$ phase remains nearly constant within pressure range of $P = 0$-8 GPa. Surprisingly enough it was found that despite $I\bar{4}3d$-Th$_4$H$_{15}$ phase was discovered five decades ago [11] there is no correct first-principles calculations reported for this phase still.

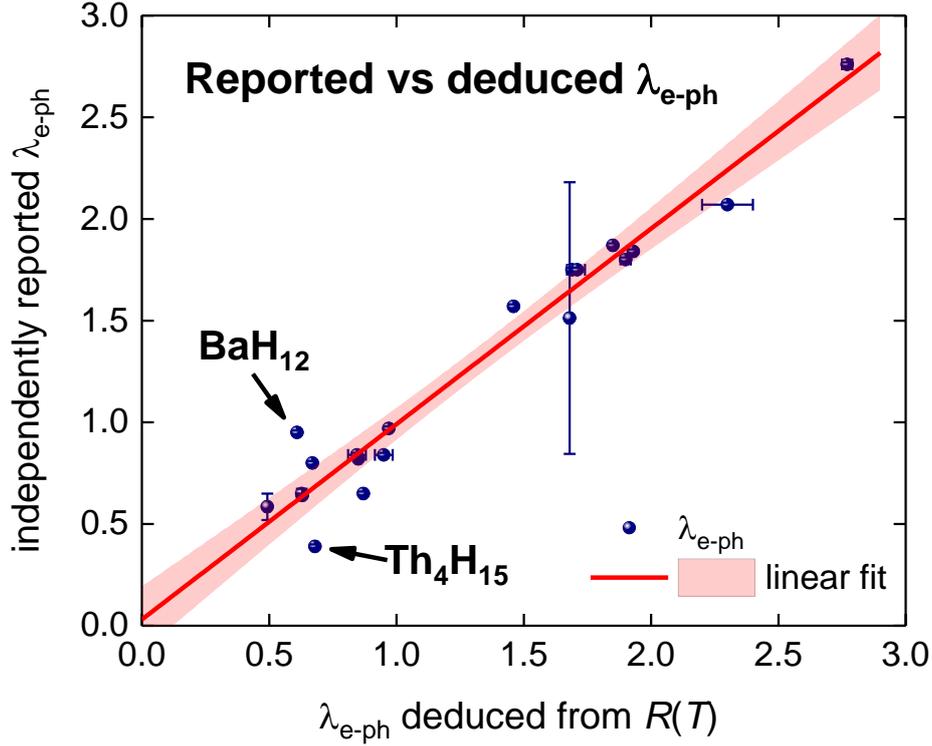

**Figure 12.** Electron-phonon coupling constant, $\lambda_{e-ph}$, deduced by Eqs. 1,3-6 vs independently reported values for the same superconductors. Deduced values for the fit of dataset to a linear equation ($y(x) = k \cdot x + b$) are: $k = 0.96 \pm 0.05$ and $b = 0.03 \pm 0.08$. Goodness of fit is 0.950. 95% confidence bands are shown by pink shadow area.

We found that there is a remarkably good agreement between $\lambda_{e-ph}$ deduced from the analysis of $R(T)$ and $\lambda_{e-ph}$ computed by first-principles calculations for highly-compressed $P6_3/mmc$-ThH$_9$ phase ($P$ = 170 GPa), $Fm\overline{3}m$-ThH$_{10}$ phase ($P$ = 170 GPa) and $Im$-$3m$-YD$_6$ phase ($P$ = 172 GPa). We also deduced Debye temperature, $T_\theta$(402 GPa) = 727 ± 6 K, and estimated the upper boundary for $\lambda_{e-ph} < 1.7$ value for metallic hydrogen phase III. These two values are in a good agreement with recent first-principles calculations [93].

**Acknowledgement**

The author thanks D. V. Semenok and A. R. Oganov (Skolkovo Institute of Science and Technology) for providing $R(T)$ data for $Fm\overline{3}m$-ThH$_{10}$ phase, and A. G. Kvashnin, I. A. Troyan, and A. R. Oganov (Skolkovo Institute of Science and Technology) for providing $R(T)$ data for *Im-3m*-YD$_6$ phase and important discussion of the results of analysis. The author thanks Jinguang Cheng (Institute of Physics, Chinese Academy of Sciences) and Jens Hänisch (Karlsruhe Institute of Technology) for detailed and fruitful discussion on the equation of the lower critical field in superconductors.

The author thanks financial support provided by the Ministry of Science and Higher Education of Russia (theme "Pressure" No. AAAA-A18-118020190104-3) and by Act 211 Government of the Russian Federation, contract No. 02.A03.21.0006.

**Data Availability Statement**

No new data were created or analysed in this study. Data sharing is not applicable to this article.